\documentclass[12pt]{article}
\usepackage{amssymb}
\usepackage{amsfonts}
\usepackage{amsmath,calrsfs}
\usepackage{bbm}

\usepackage[dvips]{epsfig}
\usepackage[utf8]{inputenc}

\def\bea{\begin{eqnarray}}
\def\eea{\end{eqnarray}}
\newcommand{\bn}{\begin{eqnarray}}
\newcommand{\en}{\end{eqnarray}}
\newcommand{\nn}{\nonumber}

\def \S {\mathbb{S}}
\def \G {\mathbb{G}}

\def \tJ {\tilde{J}}

\newcommand{\no}{\noindent}

\def\bea{\begin{eqnarray}}
\def\eea{\end{eqnarray}}

\newcommand{\be}{\begin{equation}}
\newcommand{\ee}{\end{equation}}
\def\bea{\begin{eqnarray}}
\def\eea{\end{eqnarray}}

\newcommand{\p}{\partial}
\newcommand{\hp}{\hat{\partial}}
\newcommand{\s}{\,\,\,\,}
\newcommand{\bG}{\overline{\Gamma}}
\newcommand{\bg}{\overline{\gamma}}
\newcommand{\bfi}{\overline{\phi}}
\newcommand{\tG}{\tilde{\Gamma}}
\newcommand{\tg}{\tilde{\gamma}}

\begin{document}

\title{\textbf{Soldering spin-3 opposite helicities in $D=2+1$}}
\author{D.Dalmazi$^{1}$\footnote{ddalmazi@gmail.com}, A. L. R. dos Santos$^{2}$\footnote{alessandroribeiros@yahoo.com.br},  E. L. Mendon\c ca$^{1}$\footnote{elias.leite@unesp.br}, R. Schimidt Bittencourt $^{1}$\footnote{raphael.schimidt@unesp.br} \\
    \textit{{1- UNESP - Campus de Guaratinguet\'a - DFQ} }\\
    \textit{{CEP 12516-410, Guaratinguet\'a - SP - Brazil} }\\
    \textit{{2- Instituto Tecnol\'ogico de Aeron\'autica - DCTA} }\\
    \textit{{CEP 12228-900, S\~ao Jos\'e dos Campos - SP - Brazil} }\\}
\date{\today}
\maketitle

\begin{abstract}

Here we present the ``soldering'' of opposite helicity states of a spin-3 particle, in $D=2+1$,  into one parity doublet. The starting points may be either the sixth- or the fifth-order (in derivatives) spin-3 self-dual models of opposite helicities. The high number of derivatives avoids the use of auxiliary fields which has been so far an obstacle for a successful soldering procedure. The resulting doublet model is a new Lagrangian with six orders in derivatives and no auxiliary field. It may be regarded as a spin-3 analogue of the linearized ``New Massive Gravity''. We check its particle content via a gauge invariant and Lorentz covariant analysis of the analytic structure of the two-point amplitude with the help of spin-3 analogues of the Barnes and Rivers projection operators. The particle content is alternatively confirmed in a specific non-covariant gauge by a decomposition in helicity variables. The soldered model is ghost free and contains two physical states as expected for a parity doublet.
\end{abstract}
\newpage

\section{Introduction}

Contrary to what happens in $D=3+1$ dimensions, in the lower dimension $D=2+1$ it is possible to write down  local Lagrangians for elementary spin-$s$ particles with well 
defined helicity $+s$ or $-s$. 
Those models are parity breaking (parity singlets) and may be called generically  self-dual models. Historically, the first examples correspond to the spin-1 and spin-2 cases which are known respectively as the Maxwell-Chern-Simons ($SD_2^{(1)}$)  and the linearized topologically massive gravity ($SD_3^{(2)}$) theories, see \cite{djt}. The symbol $SD_j^{(s)}$ stands for a self-dual model of helicity $s$ and of $j$-th order in derivatives. At each spin value $s=1,3/2,2$ there are $2s$ equivalent self-dual models running from the first order ($j=1$) to the top order
$j=2s$. One can go from $SD_{j-1}^{(s)}$ to $SD_{j}^{(s)}$ via a Noether gauge embedding procedure (NGE), starting with $j=2$ until $j=2s$, see \cite{clovis}, \cite{mls}, \cite{SD_4}. The more derivatives we have, the more local symmetries and the less auxiliary fields are required to get
rid of spurious degrees of freedom. This will be important for our purposes.

In the spin-3 case we have been only partially successful \cite{nge,ddhigher}. We have gone from 
$SD_1^{(3)}$ until $SD_4^{(3)}$ and from $SD_5^{(3)}$ up to the top model $SD_6^{(3)}$ along the $NGE$ and the master action approaches. We still have a gap between $SD_4^{(3)}$ and $SD_5^{(3)}$.

On the other hand, for the same set of spins $s=1,3/2,2$, one can show that opposite helicity models $SD_j^{(s)}$ 
and $SD_j^{(-s)}$ with $j=2,3, \cdots , 2s$ can be joined together into a
parity invariant (doublet) model with both helicities $\pm s$ via a 
``soldering'' procedure, see \cite{stone,bk,iw,review,gs,dm1,dm2} for references on ``soldering''. 
In particular, the spin-1 Maxwell-Proca and the spin-2 Fierz-Pauli
models can be obtained via such procedure\footnote{The linearized ``New Massive Gravity'' (NMG) of \cite{bht} can also be obtained via soldering of
linearized topologically massive gravity models of opposite helicities. The fine tuned curvature square terms $R_{\mu\nu}^2 - (3/8)R^2$ are automatically build up, at linearized level, via soldering \cite{dm1,dm2}.}  just like the spin-3/2 model of \cite{s32sold}. 
Since those doublet Lagrangians have the same form in $D=3+1$, one can regard the self-dual models in $D=2+1$ as 
building blocks of massive particles in $D=3+1$. 

It turns out that for the
next integer spin $s=3$ we have problems. The soldering procedure is more complicate  due to the presence of the auxiliary fields. In particular, we have not been able to deduce the massive spin-3 Singh-Hagen \cite{sh} model (parity doublet) completely. In \cite{tese} only the pure spin-3 sector of such model has been obtained. We
have not coped with the soldering of the auxiliary fields which are required in order to have a ghost free doublet model. Since the two highest order self-dual models $SD_6^{(3)}$ \cite{bhth} and $SD_5^{( 3)}$ \cite{ddhigher} only contain one completely symmetric rank-3 tensor without extra fields, which is the minimal tensor structure required for spin-3 particles,  they are the best candidates for the soldering procedure. The aim of this work is to show that both models can be successfully soldered into a  self-consistent doublet spin-3 model very much like the spin-2 case where a couple of opposite helicities linearized topologically massive gravities ($SD_3^{(\pm 2)}$) and linearized higher derivative topologically massive gravities ($SD_4^{(\pm 2)}$) have been both soldered into the linearized ``New Massive Gravity'' (NMG) of \cite{bht}. 

In sections 2 and 3 we solder the fifth ($SD_5^{(\pm 3)}$) and sixth ($SD_6^{(\pm 3)}$) self-dual models respectively. In section 4 we check that the sixth order soldered model is unitary in a covariant and gauge independent way. In section 5 we reaffirm the self-consistency of the doublet model in terms of helicity variables in a non covariant gauge.

\section{Soldering fifth order spin-3 self-dual models}
Along this work the spin-3 field is described in terms of totally symmetric rank-3 tensors
$h_{\mu\nu\alpha}$. There are some ``geometrical'' objects that we have named the Einstein and Schouten tensors
which are respectively given by: \bea
\mathbb{G}_{\mu\nu\alpha}=\mathbb{R}_{\mu\nu\alpha}-\frac{1}{2}\eta_{(\mu\nu}\mathbb{R}_{\alpha)}\quad,\quad\mathbb{S}_{\mu\nu\alpha}=\mathbb{R}_{\mu\nu\alpha}-\frac{1}{8}\eta_{(\mu\nu}\mathbb{R}_{\alpha)};\label{g2}\eea
\no where we have used the spin-3 Ricci tensor and its vector contraction first introduced in
\cite{deserdam}, namely:
\bea \mathbb{R}_{\mu\nu\alpha}&=&\square{h}_{\mu\nu\alpha}-\partial^{\beta}\partial_{(\mu}h_{\nu\alpha)\beta}+\partial_{(\mu}\partial_{\nu}h_{\alpha)},\label{r1}\\
\mathbb{R}_{\alpha}&=&\eta^{\mu\nu}\mathbb{R}_{\mu\nu\alpha}=2\square{h}_{\alpha}-2\partial^{\beta}\partial^{\lambda}h_{\beta\lambda\alpha}+\partial_{\alpha}\partial^{\beta}h_{\beta}. \label{r2}\eea

\no We use the mostly plus metric $(-,+,+)$ and unnormalized symmetrization:
$(\alpha\beta\gamma)=\alpha\beta\gamma+\beta\gamma\alpha+\gamma\alpha\beta$. It is useful to define the
anti-symmetric operator $E_{\mu\nu}=\epsilon_{\mu\nu\alpha}\p^{\alpha}$ where
$(Eh)_{\mu\nu\alpha}\equiv(2/3)E_{(\mu}^{\;\;\;\beta}h_{\beta\nu\alpha)}$. Given another totally symmetric
tensor $f_{\mu\nu\alpha}$, the operators $\mathbb{G}_{\mu\nu\alpha}$ and $\mathbb{S}_{\mu\nu\alpha}$ are
hermitian in the sense that under the space-time integral,

\be
\mathbb{G}_{\mu\nu\alpha}[\mathbb{S}(h)]f^{\mu\nu\alpha}=\mathbb{S}_{\mu\nu\alpha}(h)\mathbb{G}^{\mu\nu\alpha}(f)=\mathbb{S}_{\mu\nu\alpha}(f)\mathbb{G}^{\mu\nu\alpha}(h)=h_{\mu\nu\alpha}\mathbb{G}^{\mu\nu\alpha}[\mathbb{S}(f)].\ee

The fifth-order self-dual model obtained in \cite{ddhigher} describes a singlet of helicity $+3$ or $-3$ depending on the
sign in front of the highest order term \footnote{Actually the parity of the model is sensitive to the change of
$m_+ \to -m_-$.}. In this sense, let us consider: \bea
S^{(5)}_{+3}[f]=\int d^3x\Big[-\frac{1}{2m_{+}^{2}}\S_{\mu\nu\alpha}(f)\G^{\mu\nu\alpha}(f)+\frac{1}{4m^{3}_{+}}\S_{\mu\nu\alpha}(f)\G^{\mu\nu\alpha}(Ef)\Big];\label{SD5+}\eea
\bea
S^{(5)}_{-3}[g]=\int d^3x\Big[-\frac{1}{2m_{-}^{2}}\S_{\mu\nu\alpha}(g)\G^{\mu\nu\alpha}(g)-\frac{1}{4m^{3}_{-}}\S_{\mu\nu\alpha}(g)\G^{\mu\nu\alpha}(Eg)\Big].\label{SD5-}\eea

\no Where (\ref{SD5+}) represents a helicity $+3$ with mass $m_+$ and (\ref{SD5-})  a helicity $-3$ with
mass $m_-$. One can verify that they are both invariant under ``traceless reparametrizations'' and
``Weyl-transverse'' gauge transformations respectively given by:

\be \delta_{\tilde{\xi}}f_{\mu\nu\alpha}=\p_{(\mu}\tilde{\xi}_{\nu\alpha)},\label{gauge1}\ee \be
\delta_{\psi^T}f_{\mu\nu\alpha}= \eta_{(\mu\nu}\psi_{\alpha)}^T.\label{gauge2}\ee

\no where $\eta^{\mu\nu}\tilde{\xi}_{\mu\nu}=0$ and $\p^{\alpha}\psi_{\alpha}^T=0$. It is also possible to check that they are invariant under the independent global shifts: \bea
\delta{f}_{\mu\nu\alpha}=\omega_{\mu\nu\alpha}\qquad,\qquad
\delta{g}_{\mu\nu\alpha}=\kappa\omega_{\mu\nu\alpha},\eea

\no where $\omega_{\mu\nu\alpha}$ and $\kappa$ are constants. By imposing that such transformations are arbitrary space-time functions
 and proportional to each other, one can show through the soldering
procedure that the fields $f_{\mu\nu\alpha}$ and $g_{\mu\nu\alpha}$ can be tied into a gauge invariant
combination. We keep the constant $\kappa$ arbitrary so far, and then take the variations:

\bea \delta{S}^{(5)}_{+3}[f]=\int d^3x\,\,{J}^{(+)}_{\mu\nu\alpha}(f)\G^{\mu\nu\alpha}(\omega);\label{4}\eea 

\bea \delta{S}^{(5)}_{-3}[g]=\int d^3x\,\,{J}^{(-)}_{\mu\nu\alpha}(g)\G^{\mu\nu\alpha}(\kappa\omega),\label{5}\eea

\no where $J^{(+)}_{\mu\nu\alpha}$ and $J^{(-)}_{\mu\nu\alpha}$ are what we call the Noether currents defined as
\footnote{Some comments about how to determine the Noether currents in the soldering approach are given in
\cite{dm2} at the end of section 2.}:

\bea J^{(+)}_{\mu\nu\alpha}&=&-\,\frac{1}{m^{2}_{+}}\S_{\mu\nu\alpha}(f)+\frac{1}{2m^{3}_{+}}\S_{\mu\nu\alpha}(Ef)\label{J+SD5},\\
J^{(-)}_{\mu\nu\alpha}&=&-\,\frac{1}{m^{2}_{-}}\S_{\mu\nu\alpha}(g)-\frac{1}{2m^{3}_{-}}\S_{\mu\nu\alpha}(Eg).\label{J-SD5}\eea

\no By simply adding (\ref{4}) and (\ref{5}) we have: 

\bea
\delta(S^{(5)}_{+3}[f]+S^{(5)}_{-3}[g])=\int d^3x\,\,(J^{(+)}_{\mu\nu\alpha}+\kappa{J}^{(-)}_{\mu\nu\alpha})\delta{H}^{\mu\nu\alpha},\label{delta1}\eea

\no where we have introduced an auxiliary field $H_{\mu\nu\alpha}$ such that its variation is given by
$\delta{H}^{\mu\nu\alpha}=\G^{\mu\nu\alpha}(\omega)$. By rewriting the right hand side of (\ref{delta1}) with an
integration by parts, we have:
\bea
\delta\left[S^{(5)}_{+3}[f]+S^{(5)}_{-3}[g]-\int d^3x\,\,(J^{(+)}_{\mu\nu\alpha}+\kappa{J}^{(-)}_{\mu\nu\alpha})H^{\mu\nu\alpha}\right]=-\int d^3x\,\,{H}^{\mu\nu\alpha}\delta(J^{(+)}_{\mu\nu\alpha}+\kappa{J}^{(-)}_{\mu\nu\alpha}).\label{SD5+SD5}\eea

\no By explicitly calculating the currents variation one can see that they might be written as: \bea
\delta[J^{(+)}_{\mu\nu\alpha}+\kappa{J}^{(-)}_{\mu\nu\alpha}]=-\,\Big(\frac{1}{m^{2}_{+}}+\frac{\kappa^{2}}{m^{2}_{-}}\Big)\Big[\delta{H}_{\mu\nu\alpha}-\frac{1}{4}\eta_{(\mu\nu}\delta{H}_{\alpha)}\Big]+\frac{1}{2}\Big(\frac{1}{m^{3}_{+}}-\frac{\kappa^{2}}{m^{3}_{-}}\Big)\S_{\mu\nu\alpha}(E\omega).\label{Var-SD5}\eea

\no Then aiming to avoid any dynamics to the auxiliary field $H_{\mu\nu\alpha}$ one can choose the arbitrary
constant to be $\kappa^{2}=m^{3}_{-}/m^{3}_{+}$ which automatically gets rid of the last term of (\ref{Var-SD5}).
After some rearrangements we can rewrite (\ref{SD5+SD5}) as $\delta S_S=0$ where  

\bea
S_{s}=S^{(5)}_{+3}[f]+S^{(5)}_{-3}[g]-\int d^3x\Big[\frac{b}{2}H_{\mu\nu\alpha}H^{\mu\nu\alpha}-\frac{3b}{8}H_{\alpha}H^{\alpha}+H_{\mu\nu\alpha}J^{\mu\nu\alpha}\Big],\label{soldaSD5-1}\eea

\no where we have defined $J_{\mu\nu\alpha}=J_{\mu\nu\alpha}^+(f)+ \kappa J_{\mu\nu\alpha}^-(g)$ and $b=(m_{+}+m_{-})/m^{3}_{-}$. Eliminating the auxiliary field $H_{\mu\nu\alpha}$ through
its algebraic equations of motion, we finally have: \bea
S_{s}=S^{(5)}_{+3}[f]+S^{(5)}_{-3}[g]+\frac{1}{2b}\int d^3x\Big[J_{\mu\nu\alpha}J^{\mu\nu\alpha}-3J_{\mu}J^{\mu}\Big].\label{soldaSD5-2}\eea
Then, substituting back (\ref{J+SD5}) and (\ref{J-SD5}) in (\ref{soldaSD5-2}) and also defining the invariant
combination:

\be h_{\mu\nu\alpha}=\kappa{f}_{\mu\nu\alpha}-g_{\mu\nu\alpha},\ee 

\no we have the so called soldered action given by:

\bea S_{s}[h]&=&\frac{1}{c}\int d^3x\Big[\,\frac{1}{8}\S_{\mu\nu\alpha}(Eh)\G^{\mu\nu\alpha}(Eh)-\frac{(m_{+}-m_{-})}{4}\S_{\mu\nu\alpha}(h)\G^{\mu\nu\alpha}(Eh)\nn\\
&&\qquad\qquad-\,\frac{m_{+}m_{-}}{2}\S_{\mu\nu\alpha}(h)\G^{\mu\nu\alpha}(h)\Big].\label{soldaSD5-3}\eea

\no Where we have defined $c=m_-^3(m_++m_-)$. We notice that, this is a sixth order model with a fifth order
interference term proportional to the difference of masses $m_{+}-m_{-}$. It is invariant under the gauge
transformations (\ref{gauge1}) and (\ref{gauge2}) for the field $h_{\mu\nu\alpha}$. With
$m_+=m_-$ we have been able to show that in fact this model describes a doublet of
helicities $+3$ and $-3$  with no need of auxiliary fields; differently from the model (also of sixth order in
derivatives) we have obtained from the Singh-Hagen theory, through different approaches, namely the master
action \cite{hds3} and the Noether gauge embedment \cite{ngehds3}. As we will see in the next section such
result resembles the ones for the spin-2 theories.

\section{Soldering sixth order spin-3 self-dual models}
In \cite{ddhigher} the authors show that there is a master action interpolating between the fifth-order self-dual
model (\ref{SD5+}) (or (\ref{SD5-})) and a sixth-order self-dual model suggested by \cite{bhth}. One can also
verify such equivalence by means of the Noether-Gauge-Embedment approach \cite{ngehds3}.

Let us consider the spin-3 sixth-order self-dual models with different masses $m_+$  and $m_-$ respectively
given by: 

\bea
S^{(6)}_{+3}[f]=\int d^3x\Big[-\frac{1}{4m_{+}^{3}}\S_{\mu\nu\alpha}(f)\G^{\mu\nu\alpha}(Ef)+\frac{1}{8m^{4}_{+}}\S_{\mu\nu\alpha}(Ef)\G^{\mu\nu\alpha}(Ef)\Big],\label{SD6+}\eea
\bea
S^{(6)}_{-3}[g]=\int d^3x\Big[\frac{1}{4m_{-}^{3}}\S_{\mu\nu\alpha}(g)\G^{\mu\nu\alpha}(Eg)+\frac{1}{8m^{4}_{-}}\S_{\mu\nu\alpha}(Eg)\G^{\mu\nu\alpha}(Eg)\Big].\label{SD6-}\eea

\no Notice that, now the helicities $+3$ and $-3$ are determined according to the sign in front of the lowest order
term. Another difference concerns the gauge symmetries of the sixth order model. Here, (\ref{SD6+}) and
(\ref{SD6-}) are invariant under a larger set of gauge symmetries in the sense that the former traceless parameter
may now be arbitrary $\tilde{\xi}_{\nu\alpha} \to \xi_{\nu\alpha}$ in (\ref{gauge1}) as well as the transverse vector which can be
now completed with its longitudinal part $\psi_{\alpha}^T \to \psi_{\alpha}$ in (\ref{gauge2}). We begin the
soldering procedure by taking the variation of both actions and imposing that the variations of the fields
$h_{\mu\nu\alpha}$ and $f_{\mu\nu\alpha}$ are proportional to each other, exactly as we have done before in (10), then,

\bea \delta{S}^{(6)}_{+3}[f]=\int d^3x \,\,  \tJ^{(+)}_{\mu\nu\alpha}\,\,\G^{\mu\nu\alpha}(E\omega)\eea \bea
\delta{S}^{(6)}_{-3}[g]=\int d^3x \,\,  \tJ^{(-)}_{\mu\nu\alpha}\,\,\G^{\mu\nu\alpha}(\kappa E\omega);\eea

\no where in order to define the Noether currents we have factorized three derivatives through the differential operator
$\G^{\mu\nu\alpha}(E\omega)$, such that we have:

\be
\tJ^{(+)}_{\mu\nu\alpha}=\frac{1}{2m_{+}} J^{(+)}_{\mu\nu\alpha} (f)
\quad ; \quad 
\tJ^{(-)}_{\mu\nu\alpha} = \frac{1}{2m_{-}} J^{(-)}_{\mu\nu\alpha} (g) \label{J-SD6}\ee

\no So the Noether currents are exactly the same ones we had
before, except for a global factor $1/2m_{\pm}$.
After quite the same procedure one can demonstrate that we have the soldered action given by:

\bea
S_{s}=S^{(6)}_{+3}[f]+S^{(6)}_{-3}[g]-\frac{1}{2a}\int d^3x\Big[\tilde{J}_{\mu\nu\alpha}\tilde{J}^{\mu\nu\alpha}-3\tilde{J}_{\mu}\tilde{J}^{\mu}\Big],\label{soldaSD6-2}\eea

\no where we have defined as before $\tJ_{\mu\nu\alpha}= \tJ_{\mu\nu\alpha}^{(+)}(g) + \kappa \,\tJ_{\mu\nu\alpha}^{(-)}(g)$, used $\kappa=m_-^3/m_+^3$ and defined $a\equiv(m_++m_-)/4m_+m_-$. Replacing the currents (\ref{J-SD6}) in (\ref{soldaSD6-2}) and defining the invariant combination
$h_{\mu\nu\alpha}=\kappa{f}_{\mu\nu\alpha}-g_{\mu\nu\alpha}$ we obtain exactly the same doublet model we have
found in (\ref{soldaSD5-3}).

Similarities with the spin-2 case are evident at this point. In \cite{dm2} it was demonstrated
that the linearized New Massive Gravity model can be obtained through the generalized soldering of either the third ($2s-1$) or of the
fourth ($2s$) order self-dual models. This has indicated us that such model is the highest self-consistent description of a parity doublet of helicities $+2$ and $-2$. Analogously, we have seen here that the sixth order doublet model (\ref{soldaSD5-3}) is obtained by the generalized soldering of the fifth or sixth order self-dual models. Thus, we expect  (\ref{soldaSD5-3}) to be the highest spin-3 doublet model. Another reason to believe that the top order in derivatives is $2s$ again is the fact that in the master action approach, in order to derive
a dual $(j+1)$-th order model from a lower $j$-th order model it is necessary that the highest derivative term has no particle content, like a topological theory. However, the sixth order term of (\ref{soldaSD5-3}) contains a massless particle in its spectrum, as we will see in formulae (\ref{ls3}) and (\ref{tilde}) at $m\to 0$. This is exactly the same situation of the
fourth order K-term of the NMG model.


Finally, we have worked here with self-dual and doublet models of
spin-3 particles which dispense the presence of auxiliary fields and this is in fact a good reason why we could
successfully handle with the soldering approach. However, we know that another massive
spin-3 doublet model of sixth order in derivatives does exist \cite{hds3}. It contains an auxiliary scalar field besides the totally symmetric rank-3 tensor $h_{\mu\nu\alpha}$. Its sixth order term is different from the sixth order term of (\ref{soldaSD5-3}). Usually the mass term
must break the local symmetries of the kinetic (higher order) term in order to produce the so called Fierz-Pauli constraints. This is the case of our soldered action $S_s$ where the symmetry under full reparametrizations $\delta h_{\mu\nu\alpha}=\p_{(\mu}\xi_{\nu\alpha)}$ is broken down to traceless reparametrizations by the fourth order mass term. This is not the case 
of the model of \cite{hds3} where both 4th and 6th order terms are invariant only under
traceless reparametrizations.  We think that, this might be the reason
why that model requires the scalar auxiliary field. Thanks the absence of 
auxiliary fields we have been able to check here
 unitarity and particle content using the spin-projection operators displayed in 
\cite{elmrs}.

\section{Unitarity of the doublet model}
Next we show that the particle content of the sixth order model we have obtained in (\ref{soldaSD5-3}) consists of a doublet of
massive spin $+3$ and $-3$ particles in three dimensions. For sake of simplicity we now choose $m_+=m_-=m$ and
then rewrite the lagrangian in terms of spin-projection operators and
transition operators as follows: 
\bea
\mathcal{L}&=&\frac{1}{2m^4}\,\left\lbrack\frac{1}{8}\S_{\mu\nu\alpha}(Eh)\G^{\mu\nu\alpha}(Eh)-\,\frac{m^2}{2}\S_{\mu\nu\alpha}(h)\G^{\mu\nu\alpha}(h)\right\rbrack\nn\\
&=&\frac{ h_{\mu\nu\alpha}}{2}\left\lbrace \frac{\Box^3}{2m^4} P^{(3)}_{11} -
\frac{\Box^2}{2m^2}\left[ P^{(3)}_{11} + \frac{3}{8} P^{(0)}_{11} + \frac{1}{16} P^{(0)}_{22} +
\frac{\sqrt{6}}{16}\left( P^{(0)}_{12} + P^{(0)}_{21} \right)  \right]
\right\rbrace^{\mu\nu\alpha}_{\beta\lambda\sigma}  h^{\beta\lambda\sigma}.\nn\\ 
\label{lpro} \eea

We have used the same orthonormal basis of \cite{elmrs}, which is the rank three analogue of the Barnes and Rivers projection operators for rank-two tensors \cite{Barnes, Rivers}, in the sense that they are constructed from the same building blocks operators $\theta_{\mu\nu}$ and $\omega_{\mu\nu}$, for more details see our appendix. They obey the following algebra

\be P_{ij}^{(s)}P_{kl}^{(r)}=\delta^{sr}\delta_{jk}P_{il}^{(s)}.\label{algebra} \ee

\no  In our notation, the superscript $(s)$ of $P_{ij}^{(s)}$ denotes the spin subspace. If $i=j$ we have a projection operator while $i\ne j$ stands for a transition operator.
 The subscripts are used in order to count the
number of projectors of a given spin subspace, for example in the subspace of spin $0$ we have two projection operators $P_{11}^{(0)}$ and $P_{22}^{(0)}$, see (\ref{p110}) and (\ref{p220}).  In addition, the set of projectors is complete in the
sense that: \be \sum_{i,s} P^{(s)}_{ii}  =  \mathbbm{1} , \label{id} \ee

\no where $\mathbbm{1}$ stands for the  symmetric rank-3 identity operator given in (\ref{id3}).

Once the doublet model is invariant under traceless reparametrizations and Weyl transverse transformations given
respectively by (\ref{gauge1}) and (\ref{gauge2}) we need gauge fixing terms in order to obtain the propagators. In
order to fix the traceless reparametrizations, we suggest a de-Donder-like traceless tensor as
gauge condition, i.e., 

 \bea
\mathcal{L}_{GF}^{(1)} = \frac{1}{2\lambda_1} \left\lbrace  \p^\mu h_{\mu\nu\alpha} - \frac{1}{5} \left[
\p_{( \nu} h_{\alpha )} + \eta_{\nu\alpha} (\p \cdot h) \right] \right\rbrace ^2, \eea

\no where $\lambda_1$
is a gauge fixing parameter and $\p\cdot h = \p_{\mu}h^{\mu}$. We have constructed this term in such a way
that it is invariant under Weyl-transverse transformations (\ref{gauge2}). It can be rewritten  as:

 \bea
\mathcal{L}_{GF}^{(1)} &=& \frac{1}{2\lambda_1} h_{\mu\nu\alpha} \left\lbrace \Box \left[ -\frac{1}{3} P_{11}^{(2)} -\frac{8}{75} P_{11}^{(1)} -\frac{32}{75} P_{22}^{(1)} + \frac{16}{75} \left(  P_{12}^{(1)} + P_{21}^{(1)}  \right) \right]   \right\rbrace^{\mu\nu\alpha}_{\beta\lambda\sigma}  h^{\beta\lambda\sigma} \nn \\
&+& \frac{1}{2\lambda_1} h_{\mu\nu\alpha} \left\lbrace \Box \left[ -\frac{9}{25} P_{11}^{(0)} -\frac{6}{25}
P_{22}^{(0)} + \frac{3\sqrt{6}}{25} \left(  P_{12}^{(0)} + P_{21}^{(0)}  \right) \right]
\right\rbrace^{\mu\nu\alpha}_{\beta\lambda\sigma}  h^{\beta\lambda\sigma}. \eea

Since the model is still gauge invariant under Weyl-transverse transformations, we add a second gauge fixing term
given by: \bea \mathcal{L}_{GF}^{(2)} = \frac{1}{2m^6\lambda_2} f_\alpha f^\alpha \eea with: \bea f_\alpha =
\Box\tilde{f}_\alpha - \p_\alpha(\p\cdot \tilde{f}) \quad; \quad \tilde{f}_\alpha = \p^\mu\p^\nu
h_{\mu\nu\alpha} - \Box h_{\alpha}, \eea

\no which by its turn is invariant under traceless reparametrizations (\ref{gauge1}). It can be written as: \bea \mathcal{L}_{GF}^{(2)} &=& \frac{1}{2m^6\lambda_2}
h_{\mu\nu\alpha} \left\lbrace \Box^4 \left[ \frac{4}{3} P_{11}^{(1)} - \frac{1}{3} P_{22}^{(1)} \right]
\right\rbrace^{\mu\nu\alpha}_{\beta\lambda\sigma}  h^{\beta\lambda\sigma}. \eea 

\no Then considering the two gauge fixing terms, one can rewrite the lagrangian (\ref{lpro}) in a bilinear form: \bea
\mathcal{L}+\mathcal{L}_{GF}^{(1)}+\mathcal{L}_{GF}^{(2)}=h_{\mu\nu\alpha} \,\,G^{\mu\nu\alpha}_{\beta\lambda\sigma} \,\,h^{\beta\lambda\sigma} \eea
where the operator $G^{\mu\nu\alpha}_{\beta\lambda\sigma}$ can be rewritten, omitting the indices for
sake of simplicity, as:

\bea
G &=& \frac{\Box^2(\Box-m^2)}{2m^4} P_{11}^{(3)} - \frac{\Box}{3\lambda_1} P_{11}^{(2)} + \frac{4\Box}{75m^6} \left[ \frac{25\lambda_1\Box^3 - 2\lambda_2m^6}{\lambda_1\lambda_2}\right] P_{11}^{(1)} \nn \\
&-& \frac{\Box}{75m^6} \left[ \frac{25\lambda_1\Box^3 + 32\lambda_2m^6}{\lambda_1\lambda_2}\right] P_{22}^{(1)} + \frac{16\Box}{75\lambda_1} \left[ P_{12}^{(1)} + P_{21}^{(1)} \right]  \nn \\
&-& \frac{3\Box}{m^2} \left[ \frac{2\Box}{32} + \frac{3m^2}{25\lambda_1}\right] P_{11}^{(0)} -
\frac{\Box}{m^2} \left[ \frac{\Box}{32} + \frac{6m^2}{25\lambda_1} \right] P_{22}^{(0)} - \frac{\sqrt{6}\Box}{m^2}\left[ \frac{\Box}{32} - \frac{3m^2}{25\lambda_1} \right]  \left[ P_{12}^{(0)} + P_{21}^{(0)} \right] \nn \\
\eea Once we know the identity operator for symmetric rank three fields we can find the propagator: \bea
G^{-1} &=& \frac{2m^4}{\Box^2(\Box-m^2)} P_{11}^{(3)} - \frac{3\lambda_1}{\Box} P_{11}^{(2)} + \frac{3m^6}{20\Box^4} \left[ \frac{(25\lambda_1\Box^3 + 32\lambda_2m^6)\lambda_2}{5\lambda_1\Box^3+6\lambda_2m^6}\right] P_{11}^{(1)} \nn \\
&-& \frac{3m^6}{5\Box^4} \left[ \frac{(25\lambda_1\Box^3 - 2\lambda_2m^6)\lambda_2}{5\lambda_1\Box^3+6\lambda_2m^6}\right] P_{22}^{(1)} + \frac{12m^{12}}{5\Box^4}  \left[ \frac{(\lambda_2)^2}{5\lambda_1\Box^3+6\lambda_2m^6} \right]\left[  P_{12}^{(1)} + P_{21}^{(1)} \right]  \nn \\
&-& \left[ \frac{25\lambda_1\Box + 192m^2}{81\Box^2} \right] P_{11}^{(0)} - 6 \left[ \frac{25\lambda_1\Box + 48m^2}{81\Box^2} \right] P_{22}^{(0)}\nn\\
&+& \sqrt{6} \left[ \frac{25\lambda_1\Box - 96m^2}{81\Box^2} \right]  \left[ P_{12}^{(0)} + P_{21}^{(0)} \right]. \nn \\
\eea

Now in order to analyze the spectrum of the model we consider the coupling of $h_{\mu\nu\alpha}$ to the
totally symmetric source term $T^{\mu\nu\alpha}$,

 \bea S &=& \int d^3x
d^3x \left( \mathcal{L} + h_{\mu\nu\alpha} T^{\mu\nu\alpha} \right); \eea \no In order to keep the
invariance under (\ref{gauge1}) and (\ref{gauge2}) the source must satisfy the following restrictions:
\bea
\delta_{\widetilde{\xi}}S=0 &\Longrightarrow& \p_\mu T^{\mu\nu\alpha} - \frac{1}{3} \eta^{\nu\alpha} \p_\mu T^\mu = 0, \label{RF1}\\
\delta_{\psi^T}S=0 &\Longrightarrow& T^\mu = \p^\mu \varOmega. \label{RF2} \eea

\no Where $\Omega$ is an arbitrary scalar function. Now, we are ready to take the Fourier transform of the previous result in order to analyze the propagator in the momentum
space saturated by totally symmetric sources obeying the constraints (\ref{RF1}) and (\ref{RF2}). Then we look at the
imaginary part of the residue of the two point amplitude in momentum
space $\mathcal{A}_2(k)$ given by: \bea
\mathcal{A}_2(k) &=& - \frac{i}{2} T_{\mu\nu\alpha}^{*}(k)\,\, G^{-1}(k)^{\mu\nu\alpha}_{\beta\lambda\sigma} \,\,T^{\beta\lambda\sigma}(k)\\
&=& \frac{i}{k^2+m^2}\left[  T_{\mu\nu\alpha}^{*} T^{\mu\nu\alpha} - \frac{7}{9} k^2 \varOmega^2 \right] - \frac{i}{k^2}\left[  T_{\mu\nu\alpha}^{*} T^{\mu\nu\alpha} \right] \nn \\
&+& i \frac{m^2}{k^4} \left[  T_{\mu\nu\alpha}^{*} T^{\mu\nu\alpha} + k^2 \varOmega^2 \right] +
\frac{7i}{9}\varOmega^2 \label{a2k2}\eea

\no It does not depend on the gauge parameters $\lambda_1$ and $\lambda_2$. We have physical particles if $Im\left[ Res
(\mathcal{A}_2(k))\mid_{{\rm pole}} \right] > 0$.

Let us start by the massive pole analysis, which allows us to choose the convenient rest frame where
$k_\mu=(m,0,0)$. From (\ref{RF1}) and (\ref{RF2}) we have in momentum space: \be m T^{0\nu\alpha} +
\frac{i}{3} m^2 \eta^{\nu\alpha}\varOmega = 0 \ee

\no Therefore,

\bea T^{0\nu\alpha} &=& 0 \s\s\s\s\s\s\s\s\s \nu\neq\alpha \\
T^{000} &=& \frac{i}{3} m \varOmega \\
T^{0jj} &=& -\frac{i}{3} m \varOmega \s\s\s\s\s (j=1,2) \eea

\no Taking these information back in (\ref{a2k2})  we have: \bea
Im\left[ Res (\mathcal{A}_2(k))\mid_{k^2=-m^2} \right] &=& \lim\limits_{k^2 \xrightarrow{} -m^2} (k^2+m^2) \mathcal{A}_2(k) \nn \\
&=& |T_{ijk}|^2 >0 \s\s\s (i,j,k=1,2) \eea

Hence, a physical massive spin-3 particle is propagating in the spectrum. However we still have a double massless pole
in the spin-3 sector of $G^{-1}$ which deserves special care. In order to
analyze it we choose the frame $k_\mu=(-k_0,\epsilon,-k_0)$ which implies $k^2=\epsilon^2$. At the end we take the limit $\epsilon\longrightarrow 0$.  From the constraints (\ref{RF1}) and (\ref{RF2})
we can eliminate 7 of the 10 independent components of the totally symmetric source, in such a way that we
can conveniently choose as independent variables $\Omega$, $T^{122}$ and $T^{022}$. Exactly as in the analysis carried out 
in \cite{ddunitarity} other choices may require specific properties of some of the components of $T_{\mu\nu\alpha}$
at $\epsilon \to 0$ in order to guarantee that all $T_{\mu\nu\alpha}$ behave smoothly at such limit. Explicitly we have,

\bea
T^{000}&=& i\varOmega \frac{2\epsilon (\epsilon/k_0) \left[  5 - 3(\epsilon/k_0)^2 \right]  }{3 \left[ 1 - (\epsilon/k_0)^2 \right]^2 } + T^{022} + \frac{2(\epsilon/k_0)\left[ 1 + (\epsilon/k_0)^2 \right] }{\left[ 1 - (\epsilon/k_0)^2\right]^2 }T^{122} \\
T^{001}&=& i\varOmega \frac{2\epsilon \left[  3 - (\epsilon/k_0)^4  \right]  }{3 \left[ 1 - (\epsilon/k_0)^2 \right]^2 } + \frac{\left[ 1 + (\epsilon/k_0)^2 \right]^2 }{\left[ 1 - (\epsilon/k_0)^2\right]^2 }T^{122} \\
T^{002}&=& i\varOmega \frac{\epsilon (\epsilon/k_0) \left[ -3 + (\epsilon/k_0)^2 \right]}{ 3 \left[ 1 - (\epsilon/k_0)^2 \right] } -T^{022} - \frac{(\epsilon/k_0)\left[ 1 + (\epsilon/k_0)^2 \right] }{\left[ 1 - (\epsilon/k_0)^2\right] }T^{122} \\
T^{011}&=& i\varOmega \frac{k_0\left[ 3 + 4(\epsilon/k_0)^2 - 3(\epsilon/k_0)^4\right] }{3 \left[ 1-(\epsilon/k_0)^2 \right]^2 } + \frac{2(e/k_0)\left[ 1 + (\epsilon/k_0)^2 \right] }{\left[ 1-(\epsilon/k_0)^2 \right]^2}T^{122} \\
T^{012}&=& i\varOmega \frac{\epsilon\left[ -3 + (\epsilon/k_0)^2\right] }{3\left[ 1 - (\epsilon/k_0)^2\right] } - \frac{\left[ 1 + (\epsilon/k_0)^2 \right] }{\left[ 1 - (\epsilon/k_0)^2 \right]}T^{122} \\
T^{111}&=& i\varOmega \frac{\epsilon\left[ 9 - 6(\epsilon/k_0)^2 + (\epsilon/k_0)^4\right] }{3\left[ 1 - (\epsilon/k_0)^2\right]^2 } + \frac{4(\epsilon/k_0)^2 }{\left[ 1 - (\epsilon/k_0)^2\right]^2}T^{122} \\
T^{112}&=& i\varOmega \frac{k_0\left[ -3 + (\epsilon/k_0)^2\right] }{3\left[ 1- (\epsilon/k_0)^2\right] } - \frac{2(\epsilon/k_0)}{\left[ 1- (\epsilon/k_0)^2\right]}T^{122} \\
T^{222}&=&-i\varOmega \frac{\epsilon(\epsilon/k_0)}{3} - T^{022} + (\epsilon/k_0)T^{122} \eea

Collecting all the previous results we can write : \bea
T_{\mu\nu\alpha}^{*}T^{\mu\nu\alpha}&=&-|T^{000}|^2+3|T^{001}|^2+3|T^{002}|^2-3|T^{011}|^2-6|T^{012}|^2 \nn \\
& & -3|T^{022}|^2+|T^{111}|^2+3|T^{112}|^2+3|T^{122}|^2+|T^{222}|^2\\ \\
&=&-\frac{(k_0)^2(\epsilon/k_0)^2\left[4(\epsilon/k_0)^4 - 9(\epsilon/k_0)^2 + 9 \right]}{9\left[ 1 - (\epsilon/k_0)^2 \right] } \varOmega^2 \nn \\
& & + \frac{4}{3} \frac{\epsilon (\epsilon/k_0)^3\left[(\epsilon/k_0)^2 - 2\right] }{\left[ 1 - (\epsilon/k_0)^2\right]^2 }\left\lbrace i\varOmega T^{*022} - i\varOmega^* T^{022} \right\rbrace \nn \\
& & -\frac{4}{3} \frac{k_0 (\epsilon/k_0)^5\left[(\epsilon/k_0)^2 - 2 \right] }{\left[ 1 - (\epsilon/k_0)^2\right]^2 }\left\lbrace i\varOmega T^{*122} - i\varOmega^* T^{122} \right\rbrace \nn \\
& & -\frac{4 (\epsilon/k_0)^5}{\left[ 1 - (\epsilon/k_0)^2\right]^2 }\left\lbrace T^{022}T^{*122} + T^{*022} T^{122} \right\rbrace \nn \\
& & +\frac{4 (\epsilon/k_0)^4 \left[ 1 + (\epsilon/k_0)^2 \right] }{\left[ 1 - (\epsilon/k_0)^2\right]^2
}|T^{122}|^2 \eea which reduces to the simple expression: \bea T_{\mu\nu\alpha}^{*}T^{\mu\nu\alpha}&\approx&
-\epsilon^2\varOmega^2 + {\cal O}(\epsilon^3), \eea

 Then we have: \bea
Im\left[ Res (\mathcal{A}_2(k))\mid_{k^2=0}\right] &=& \lim\limits_{\epsilon \longrightarrow 0} \epsilon^2 \mathcal{A}_2(k) \nn \\
&=& \lim\limits_{\epsilon \longrightarrow 0} \left\lbrace  \epsilon^2\varOmega^2 - O(\epsilon^3) + \frac{m^2}{\epsilon^2} \left[ O(\epsilon^3) \right] \right\rbrace=0 \nn \\
\eea We finally verify that the massless pole is non propagating. After all, we conclude that the higher
derivative massive spin-3 doublet model is free of ghosts and carries only one massive spin $+3$
particle (parity doublet) in $D=2+1$ dimensions.

\section{Particle content via helicity variables}

Since the particle content analysis of last section 
is rather technical, we present here an alternative analysis based on the less technical, though not explicitly covariant,  approach of \cite{deserprl}, see also  \cite{andringa,bhth} and 
more recently \cite{ddhigher}. They make use of helicity variables and convenient gauge conditions fixed at action level. Our starting point is the soldered action (\ref{soldaSD5-3}) which at $m_+ = m_- = m$  becomes:

\be S_{s}[h]=\frac{1}{4\, m^4}\int d^3x {\cal L}_s = \frac{1}{4\, m^4}\int d^3x \left\lbrack {\cal L}_{(6)} - m^2\, {\cal L}_{(4)}\right\rbrack 
\label{ls} \ee

\no The sixth and fourth order Lagrangians are given by

\bea {\cal L}_{(6)} &=& \frac 14 \S_{\mu\nu\alpha}(h)\G^{\mu\nu\alpha}(E^2 \, h) = h_{\mu\nu\sigma} \Box^3\left(
\theta^{\mu\alpha}\theta^{\nu\beta} - \frac 34 \theta^{\mu\nu}\theta^{\alpha\beta}\right) \theta^{\sigma\lambda}
h_{\alpha\beta\lambda} \label{l6a}\\ {\cal L}_{(4)} &=& \S_{\mu\nu\alpha}(h)\G^{\mu\nu\alpha}( h ) \label{l4a} \eea

The reader can check that both (\ref{l6a}) and (\ref{l4a}) are invariant under traceless reparametrizations and
transverse Weyl transformations, see (\ref{gauge1}) and (\ref{gauge2}). In total we have seven independent gauge
parameters among  $\tilde{\xi}_{\nu\alpha}$ and $\psi_{\mu}^T$ which allow us to fix seven gauge conditions.
Initially we fix the same five gauge conditions used in \cite{ddhigher} since they are rather convenient,
namely,

\be \p_{j} h_{jk\mu} = 0 \quad , \quad  \, j,k = 1,2 \,\, ; \, \,  \mu = 0,1,2 \label{gc1} \ee

\no According to \cite{moto1} we can safely fix gauge conditions at action level if they are {\it
complete}. In our case this means that the five gauge conditions (\ref{gc1}) must {\it completely} fix (without
ambiguity) five out of the seven independent gauge parameters $(\tilde{\xi}_{\nu\alpha},\psi_{\mu}^T)$. As shown in
\cite{ddhigher}, the conditions (\ref{gc1}) do satisfy such criterium. We can further fix the two remaining
gauge degrees of freedom. However, we need to be careful in order to preserve the completeness property of all
seven gauge conditions simultaneously. If we apply the gauge transformations (\ref{gauge1}) and (\ref{gauge2})
on (\ref{gc1}) and look for residual symmetries which leave it invariant, we completely determine the five
parameters $\tilde{\xi}_{\nu\alpha} $ as functions of the two independent Weyl parameters contained in
$\psi_{\mu}^T$. Then, we can select combinations of the fields $h_{\mu\nu\alpha}$ and its derivatives which are
pure gauge under such residual symmetries. Such combinations can be used as complete gauge conditions. Following
that route we end up with the two remaining conditions:

\be \hp_j \hp_k \hp_l h_{jkl} = 0  \quad ; \quad \nabla^2 h_{000} - 6\, \hp_j\hp_k h_{jk0} = 0 \label{gc2} \ee

\no where $\hat{\p}_j = \epsilon_{jk}\p_k $ satisfies $\hat{\p}_i\hat{\p}_j = \nabla^2 \delta_{ij} - \p_i\p_j$
and $\hat{\p}_i\hat{\p}_i = \p_j\p_j = \nabla^2$. The general solution\footnote{In \cite{ddhigher} we have only fixed (\ref{gc1}) but we could have fixed (\ref{gc2}) too which would have saved some steps in the proof of absence of particle content of ${\cal L}_{(4)}$.}, see \cite{ddhigher}, to (\ref{gc1}) and (\ref{gc2}) can be written in terms of three fields. Following the notation of \cite{ddhigher} we write:

\bea h_{jkl} &=& 0 \quad ; \quad h_{jk0} = \hat{\p}_j  \hat{\p}_k \phi \quad , \label{phi} \\
h_{00j} &=& \hat{\p}_j  \gamma + \p_j \Gamma \quad ; \quad h_{000} = 6 \nabla^2 \phi \quad , \label{rho} \eea

\no Back in the soldered theory (\ref{soldaSD5-3}) we can write, after integrations by parts, the soldered
Lagrangian ${\cal L}_s = {\cal L}_{(6)} - m^2 {\cal L}_{(4)}$ as follows

\be {\cal L}_s =  \frac{81}{4} \bfi \nabla^8(m^2-\nabla^2)\bfi + \frac{27}4 \bfi \nabla^8 \dot{\bG} - \frac
9{16} \dot{\bG} \nabla^6 \dot{\bG} + \frac 9{16} m^2\, \bG \nabla^6 \bG - \frac 9{4}  \bg \nabla^6 (\Box -
m^2)\bg \\ \label{ls2} \ee

\no where we have used the same field redefinitions of \cite{ddhigher}, i.e.,

\be \bfi = \phi - \frac{\ddot{\phi}}{6\nabla^2} - \frac{\dot{\Gamma}}{6\nabla^2} \quad ; \quad \bG = \Gamma +
\dot{\phi} \quad \label{bar} \ee

\no Although (\ref{bar}) contain time derivatives, the Jacobian is trivial ($J=1$) and the canonical structure
of the theory is preserved. We can freely invert $(\phi,\Gamma)$ in terms of $(\bfi,\bG)$. After another round
of canonically trivial redefinitions we can finally write the soldered theory in a diagonal form:

\be {\cal L}_s = \tG(\Box - m^2)\tG + \tg(\Box - m^2)\tg + \tilde{\phi} \nabla^8(m^2-\nabla^2)\tilde{\phi}
\label{ls3} \ee

\no where

\be \tG = \frac{3\, m}4\left(\frac{-\nabla^6}{m^2-\nabla^2}\right)^{1/2} \bG \quad ; \quad \tg = \frac 32
(-\nabla^2)^{3/2}\gamma \quad ; \quad \tilde{\phi} = \bfi + \frac{\dot{\bG}}{6(m^2-\nabla^2)} \quad
\label{tilde} \ee

\no Since the eigenvalues of $-\nabla^2$ are definite
positive, we can go back to our original fields $(\phi,\gamma,\Gamma)$  without problems. 

The last term in (\ref{ls3}) shows that $\tilde{\phi}$ is non propagating. Thus, we end up with only two propagating physical degrees of freedom  $(\tG,\tg)$ with the same mass, corresponding to the $+3$ and $-3$ helicity states which confirms the spectrum obtained in the last section via the analytic structure of the propagator. The approach used here can implemented in the more general case with $m_+ \ne m_-$.

As a last remark we notice that the soldered Lagrangian acquires a quite simple form in terms of spin-3 Ricci-like \cite{deserdam} curvatures:

\be {\cal L}_s = \mathbb{R}_{\mu\nu\alpha} (\Box - m^2) \mathbb{R}^{\mu\nu\alpha} - \frac{15}{16} \mathbb{R}_{\mu} (\Box - m^2) \mathbb{R}^{\mu} +\frac{(\p_{\mu}\mathbb{R}^{\mu})^2}{16} \label{ricci} \ee

\no The relative factor $-15/16$ guarantees that the first two terms
proportional to the Klein-Gordon operator are invariant under transverse Weyl transformations $\delta h_{\mu\nu\rho} = \eta_{(\mu\nu }\psi_{\rho )}^T $ under which the last term of (\ref{ricci}) is automatically invariant. The last term is however, necessary to make the sixth order terms (mass independent ones) invariant under full Weyl transformations where $\psi_{\mu}^T \to \psi_{\mu} $. It is usually necessary in massive spinning particles that the mass term breaks local symmetries of the highest derivative term in order to produce the Fierz-Pauli conditions required to achieve the correct number of degrees of freedom like in Maxwell-Proca theory. The mass terms in (\ref{ricci}) break exactly one degree of freedom of symmetry just like the Einstein-Hilbert term breaks the scalar Weyl symmetry ($\delta h_{\mu\nu}=\eta_{\mu\nu} \phi $) of the fourth order K-term of the ``New Massive Gravity'' \cite{bht}.

\section{Conclusion}

In $D=2+1$ we can solder opposite helicities theories (self-dual models) into local field
theories describing usual massive spinning particles. Thus, we can regard the self-dual models (parity singlets) as the basic building blocks of massive spinning particles (parity doublets). The soldering procedure has been successfully applied for particles of spin $s=1,3/2,2$. However, when we try to extend this idea to spin-3 particles, due to the auxiliary fields, we have only partial success. Here we have surmounted this problem by making use of higher order self-dual models described solely in terms of 
totally symmetric rank-3 tensors $h_{\mu\nu\rho}$
 which is the minimal tensor structure required for spin-3.
This is the first successful soldering beyond $s=2$ and the soldered theory (\ref{ls}) is the first spin-3 parity doublet with the minimal tensor structure. The price we have paid is to end up with six derivatives in the model, see (\ref{ricci}) and (\ref{r1}), (\ref{r2}). 

Although we have higher derivatives we have shown in section 4 that the model is unitary via a careful examination of the analytic structure of two point amplitude. The proof is Lorentz covariant and gauge independent. In section 5, by means of helicity variables,  we have reaffirmed the results of section 4 in a less technical way in a non covariant gauge. We have shown that the theory contains only two physical massive modes in the spectrum.

It is important to mention that a successfull soldering of spin-3 particles is quite unexpected from the point of view of a possible spin-3 geometry, see comment \cite{spin4}. Though we still do not know what is the natural (if any) higher spin analogue of the spin-2 Einstein tensor, Schouten tensor, etc, it seems reasonable to define in $D=2+1$, see \cite{bhth} and \cite{hhl},  a spin-s Einstein tensor of s-th order in derivatives: $G_{\mu_1\mu_2 \cdots \mu_s}= E_{\mu_1}^{\,\,\nu_1}\cdots E_{\mu_s}^{\,\,\nu_s}h_{\nu_1 \cdots \nu_s}$, where $E^{\mu\nu} = \epsilon^{\mu\nu\rho}\p_{\rho}$. Accordingly, for spin-3 we would have a third order Einstein tensor which differs from the second order one given in (\ref{g2}) which on its turn follows from the spin-3 geometry suggested in \cite{deserdam}. In the spin-2 case both definitions coincide which makes the spin-3 case rather interesting. 

 Starting with a third order spin-3 Einstein tensor the authors of \cite{bhth} suggest a fifth-order analogue of the spin-2 ``New Massive Gravity'' (NMG) of \cite{bht}. It turns out that such model contais two degrees of freedom one of which is a ghost. Since the NMG theory can be obtained from the soldering of two  linearized topologically massive gravities with opposite helicities, \cite{dm1} this makes the spin-3 soldered version of NMG unlikely as mentioned in \cite{spin4}. According to our results, one might also  consider the soldered Lagrangian (\ref{ricci}) a spin-3 analogue of the NMG model, since it is of order $2s$ and stems from the soldering of the 
opposite helicity self-dual models of order $2s$ or $2s-1$. Moreover, the local symmetry of the sixth ($2s$) order terms of (\ref{ricci}) differ from the symmetries of the fourth ($2s-2$) order terms (mass terms) by exactly one degree of freedom just like the case of the NMG model.
Moreover, when written in terms of spin projection operators, the sixth order term of $S_S$ only belongs to the spin-3 subspace just like the NMG fourth order term lies completely in the spin-2 sector.

The difference between the third and the second order spin-3 Einstein tensors is related to the choice of full reparametrizations $\delta h_{\mu\nu\rho} = \p_{(\mu}\Lambda_{\nu\rho)}$ or traceless reparametrizations $\delta h_{\mu\nu\rho} = \p_{(\mu}\tilde{\Lambda}_{\nu\rho)}$ 
respectively as the spin-3 analogue of the linearized general coordinate invariance $\delta h_{\mu\nu} = \p_{\mu}\Lambda_{\nu} + \p_{\nu}\Lambda_{\mu} $. The simplicity of our soldered action (\ref{ricci}) when written in terms of the Ricci-like curvature (\ref{r1}) invariant under traceless reparametrizations seems to favour the second choice but we have no definite conclusion about it. 

Our results raise some interesting points to be investigated in the future. It is known \cite{jm} that the fourth order NMG model can be obtained from an unconventional dimensional reduction of the second order linearized Einstein-Hilbert massless theory, we are currently investigating the possibility of deriving the soldered model
(\ref{ricci}) from the massless Fronsdal \cite{fronsdal} spin-3 model.
This is somehow awkward since dimensional reduction of massless theories with
restricted (traceless) symmetries usually leads to more fields than we originally have, 
however both (\ref{ricci}) and the spin-3 Fronsdal theories only depend on the
totally symmetric rank-3 field. Another interesting point is the possible generalization to the spin-4 case where our results could be related with the sixth order ghost free doublet model
obtained in \cite{spin4}. Finally, we mention the possibility of investigating possible cubic vertices to be added to the soldered action in order to preserve its local symmetries and derive a self consistent self-interacting spin-3 model.

\section{Appendix}

Taking the spin-1 and spin-0 projection operators $\theta_{\mu\nu}=\eta_{\mu\nu}-\omega_{\mu\nu}$ and
$\omega_{\mu\nu}={\p_{\mu}\p_{\nu}}/{\Box}$, one can construct in $D$ dimensions, the spin-3 projection
operators as follows:

\bea
(P^{(3)}_{11})^{\mu\nu\rho}_{\alpha\beta\gamma} &=&  \theta^{(\mu}_{(\alpha} \theta^\nu_\beta \theta^{\rho)}_{\gamma)} - (P^{(1)}_{11})^{\mu\nu\rho}_{\alpha\beta\gamma},\label{first} \\
(P^{(2)}_{11})^{\mu\nu\rho}_{\alpha\beta\gamma} &=&  3 \theta^{(\mu}_{(\alpha} \theta^\nu_\beta \omega^{\rho)}_{\gamma)} - (P^{(0)}_{11})^{\mu\nu\rho}_{\alpha\beta\gamma},  \\
(P^{(1)}_{11})^{\mu\nu\rho}_{\alpha\beta\gamma} &=&  \frac{3}{(D+1)}  \theta^{(\mu\nu} \theta_{(\alpha\beta}  \theta^{\rho)}_{\gamma)},\\
(P^{(1)}_{22})^{\mu\nu\rho}_{\alpha\beta\gamma} &=&  3 \theta^{(\mu}_{(\alpha} \omega^\nu_\beta \omega^{\rho)}_{\gamma)},  \\
(P^{(0)}_{11})^{\mu\nu\rho}_{\alpha\beta\gamma} &=&  \frac{3}{(D-1)}  \theta^{(\mu\nu} \theta_{(\alpha\beta}  \omega^{\rho)}_{\gamma)}, \label{p110} \\
(P^{(0)}_{22})^{\mu\nu\rho}_{\alpha\beta\gamma} &=&   \omega_{\alpha\beta}
\omega^{\mu\nu} \omega^{\rho}_{\gamma} \label{p220} \eea

\no We emphasize that here, differently from section-2, the parenthesis means normalized symmetrization, taking for example the first term in (\ref{first}) we have: \be \theta^{(\mu}_{(\alpha} \theta^\nu_\beta \theta^{\rho)}_{\gamma)}=\frac{1}{6}(
\theta^{\mu}_{\alpha} \theta^\nu_\beta \theta^{\rho}_{\gamma}+
\theta^{\rho}_{\alpha} \theta^\nu_\beta \theta^{\mu}_{\gamma}+
\theta^{\nu}_{\alpha} \theta^\mu_\beta \theta^{\rho}_{\gamma}+
\theta^{\rho}_{\alpha} \theta^\mu_\beta \theta^{\nu}_{\gamma}+
\theta^{\nu}_{\alpha} \theta^\rho_\beta \theta^{\mu}_{\gamma}+
\theta^{\mu}_{\alpha} \theta^\rho_\beta \theta^{\nu}_{\gamma}).\ee   
The totally symmetric identity operator is represented by $\mathbbm{1}$
and is given by: \bea \mathbbm{1}^{\mu\nu\rho}_{\alpha\beta\gamma} = 
\delta^{(\mu}_{(\alpha} \delta^\nu_\beta \delta^{\rho)}_{\gamma)}. \label{id3} \eea Finally, the transition operators $P^{(s)}_{{ij}}$
are given by: \bea
(P^{(1)}_{{12}})^{\mu\nu\rho}_{\alpha\beta\gamma} &=&  \frac{3}{\sqrt{(D+1)}}   \theta_{(\alpha\beta} \theta^{(\rho}_{\gamma)} \omega^{\mu\nu)},  \\
(P^{(1)}_{{21}})^{\mu\nu\rho}_{\alpha\beta\gamma} &=&  \frac{3}{\sqrt{(D+1)}}   \theta^{(\mu\nu} \theta^{\rho)}_{(\gamma} \omega_{\alpha\beta)},  \\
(P^{(0)}_{{12}})^{\mu\nu\rho}_{\alpha\beta\gamma} &=&  \frac{3}{\sqrt{3(D-1)}}   \theta_{(\alpha\beta} \omega^{(\mu\nu} \omega^{\rho)}_{\gamma)},   \\
(P^{(0)}_{{21}})^{\mu\nu\rho}_{\alpha\beta\gamma} &=& 
\frac{3}{\sqrt{3(D-1)}}   \theta^{(\mu\nu} \omega_{(\alpha\beta} \omega^{\rho)}_{\gamma)}. \eea \no

\section{Acknowledgements}

D.D. is partially supported by CNPq  (grant 306380/2017-0), the work of A.L.R.dos S. has been supported by CNPq-PDJ (grant 150524/2018-8),
while the work of R.S.B has been suported by CAPES.

\end{document}